\documentclass[aps,pra,showpacs,twocolumn]{revtex4}
\usepackage{amsfonts,amssymb,graphicx}
\usepackage[usenames,dvipsnames]{xcolor}
\usepackage{pdfpages}
\usepackage{footmisc}
\usepackage[normalem]{ulem}
\usepackage{color}

\begin{document}

\title{Vibration-assisted exciton transfer in molecular aggregates strongly coupled to confined light fields}
\author{Jingyu Liu}
\affiliation{Center for Quantum Technology Research, School of Physics, Beijing Institute of Technology, Beijing 100081, China}
\author{Qing Zhao}
\affiliation{Center for Quantum Technology Research, School of Physics, Beijing Institute of Technology, Beijing 100081, China}
\author{Ning Wu}
\email{wun1985@gmail.com}
\affiliation{Center for Quantum Technology Research, School of Physics, Beijing Institute of Technology, Beijing 100081, China}

\begin{abstract}
We investigate exciton transport through one-dimensional molecular aggregates interacting strongly with a cavity mode. Unlike several prior theoretical studies treating the monomers as simple two-level systems, exciton-vibration coupling is explicitly included in the description of open quantum dynamics of the system. In the framework of the Holstein-Tavis-Cummings model with truncated vibrational space, we investigate the steady-state exciton transfer through both a molecular dimer and longer molecular chains. For a molecular dimer, we find that vibration-assisted exciton transfer occurs at strong exciton-cavity coupling regime where the vacuum Rabi splitting matches the frequency of a single vibrational quanta. Whereas for longer molecule chains, vibration-assisted transfer is found to occur at the ultrastrong exciton-cavity coupling limit. In addition, finite relaxation of vibrational modes induced by the continuous phonon bath is found to further facilitate the exciton transport in vibrational enhancement regimes.
\end{abstract}


\maketitle

\section{Introduction}
\par Excitation energy transfer is a process of fundamental importance in a variety of physical phenomena, such as natural photosynthesis~\cite{Photosy1,Photosy2}, organic semiconductor and solar cell physics~\cite{orgsem1,orgsem2}, and artificial quantum simulators~\cite{PRX}, among others. In these systems, the excitation energy are usually carried by excitons that are bound electron-hole pairs. Recently, there have been extensive study, both experimentally and theoretically, on exciton transfer in molecular systems under the influence of strong coupling with confined light fields~\cite{Natm2015,FJ2015,enhanced2015,Zhou2016,enhanced2017,Zhong,Zhou2018,FJ2018,SR2018}. It is revealed in these works that exciton-type transport in organic materials can be dramatically enhanced when the molecules are strongly coupled to certain confined light fields, such as electromagnetic modes in metallic microcavities and/or surface plasmonic modes. To explain this intriguing phenomenon, it is argued that strong coupling between excitonic and photonic degrees of freedom leads to the formation of delocalized polaritonic modes, which provide an efficient channel for localized excitations to bypass the molecular array~\cite{Natm2015,FJ2015,enhanced2015}.
\par Organic materials offer an excellent platform to enter the strong light-matter interaction regime due to their large dipole moments and high achievable molecular densities~\cite{Mazza}. In practice, the exciton transfer efficiency generally depends on a wide range of factors, among which fluctuations caused by intramolecular vibrations and surrounding environment play a dominant role. However, simulating the transport dynamics in the presence of the molecular environment requires solving fully quantized models, which is a difficult task for molecular systems of large sizes. Thus, in most theoretical descriptions of cavity-assisted exciton transport, the molecular systems are often approximately modeled as simple two-level systems, with effects of the intramolecular vibrations being ignored~\cite{Natm2015,FJ2015,enhanced2015}.
\par Recently, there have appeared several theoretical works that explicitly include intramolecular vibrations in the original exciton-cavity system~\cite{Ephy,Spano2015,Spano2016,FJPRX,LPP,KeelingACS}. The resultant composite system, which involves excitonic, photonic, and vibrational degrees of freedom, can be described by the so-called Holstein-Tavis-Cummings model. In the framework of the Holstein-Tavis-Cummings model, a variety of static properties of organic materials strongly coupled to light fields are revealed. Examples include vibrational decoupling of dark excitons~\cite{Spano2015,LPP} and enhancement of vibrational dressing of the cavity mode~\cite{LPP} in the strong coupling regime, calculation of spectra of vibrationally dressed polaritons~\cite{KeelingACS}, and so on. Although both photonic~\cite{Natm2015,FJ2015,enhanced2015,enhanced2017,Zhong} and vibrational effects~\cite{PRX,PRB2005,PRB2011,Natphy2013,NC2014,Plenio2015} on exciton/energy transport in several natural or artificial physical systems have been investigated extensively in recent years, the interplay between the two has been much less studied.
\par In this work, in the framework of the Holstein-Tavis-Cummings model, we study combinational effects of exciton-photon and exciton-vibration coupling on exciton transfer through one-dimensional molecular aggregates. Working in truncated vibrational spaces with fixed total numbers of vibrations, we calculate the exciton transfer efficiency through a source-molecular bridge-drain setup by simulating the open dynamics of the hybrid system involving excitonc, photonic, and vibrational degrees of freedom. For small aggregates such as a molecular dimer, it is found that increase of the exciton-vibration coupling strength generally suppresses the exciton transfer efficiency in both the weak- and ultrastrong-coupling limits. However, vibration-assisted transfer is observed for strong exciton-cavity couplings at which the resulting vacuum Rabi splitting matches the energy of a single vibrational quanta. For longer molecular chains with $N\geq 4$ monomers, we observe vibration-enhanced exciton transfer in the ultrastrong exciton-cavity coupling limit. Furthermore, finite relaxation of vibrational modes are always found to facilitate the exciton transfer when vibration-enhanced exciton transfer occurs.

\par The rest of the paper is organized as follows. In Sec.~\ref{SecII}, we introduce our model and the master equation that describes the open dynamics of the hybrid system. In Sec.~\ref{SecIII}, we study exciton transport in the absence of vibrations and compare the obtained results with previous literatures. In Sec.~\ref{SecIV} we study exciton transport in the presence of vibrations for different sizes of molecular aggregates. Conclusions are drawn in Sec.~\ref{SecV}.
\section{The setup}\label{SecII}
\par To describe exciton transport through a one-dimensional molecular aggregate, we consider a source-bridge-drain setup (see Fig.~\ref{Fig1}) analogous to that used to treat electron transport in semiconductor quantum dots~\cite{Datta}. Similar approach has also been used in the study of exciton currents in photosynthetic systems~\cite{Guan2013}.
\par A one-dimensional molecular aggregate with $N$ monomers (with free ends) can be described by the Holstein model~\cite{LPP}(setting $\hbar=1$)
\begin{eqnarray}\label{H_hol}
H_{\rm{Hol}}&=&H_{\rm{e}}+H_{\rm{v}}+H_{\rm{e-v}},\nonumber\\
H_{\rm{e}}&=&\sum^N_{n=1}\varepsilon_n a^\dag_n a_n +J\sum^{N-1}_{n=1}(a^\dag_na_{n+1}+a^\dag_{n+1}a_n),\nonumber\\
H_{\rm{v}}&=&\omega_0\sum^N_{n=1}b^\dag_n b_n,\nonumber\\
H_{\rm{e-v}}&=&\lambda\omega_0\sum^N_{n=1}a^\dag_n a_n(b_n+b^\dag_n).
\end{eqnarray}
where $a^\dag_n$ creates an excitonic state $|n\rangle_{\mathrm{e}}$ on site $n$ with excitation energy $\varepsilon_n$, and $J$ is the uniform exciton hopping integral between nearest-neighbor monomers. The boson operator $b^\dag_n$ creates an intramolecular vibration on site $n$ with uniform frequency $\omega_0$. $H_{\rm{e-v}}$ is the linear exciton-vibration coupling with strength measured by the Huang-Rhys factor $\lambda^2$.
\par When the molecule is located in a single-mode microcavity described by
\begin{eqnarray}\label{H_c}
H_{\rm{c}}=\omega_{\rm{c}}c^\dag c,
\end{eqnarray}
with photon creation operator $c^\dag$ and cavity frequency $\omega_{\rm{c}}$, the exciton-cavity interaction has the form
\begin{eqnarray}\label{H_ec}
H_{\rm{e-c}}=g\sum_n(a^\dag_n c+c^\dag a_n),
\end{eqnarray}
where $g$ is the uniform exciton-cavity interaction strength, which is a good approximation when the sizes of the aggregates are smaller compared with the optical wavelength. Here, we have used the rotating wave approximation, which is also a good approximation as long as the ultrastrong light-matter coupling with $g\sqrt{N}\geq\varepsilon_n,\omega_{\rm{c}}$ is not reached. The effects of counter-rotating terms on excitonic spectral features in strongly-coupled organic molecules were discussed in Ref.~\cite{KeelingnonRWA} by treating these terms as perturbations.
\par We assume that the first (last) monomer connects to an exciton source (drain), which is analogous to the left (right) lead in a typical electron transport setup. We also assume weak interaction between the exciton reservoirs and the monomers, so that we can follow standard Born-Markov approximation to get the following dissipators in the master equation for the reduced density matrix $\rho$ of the hybrid system (i.e., the exciton-cavity-vibration system):
\begin{figure}
\includegraphics[width=0.5\textwidth]{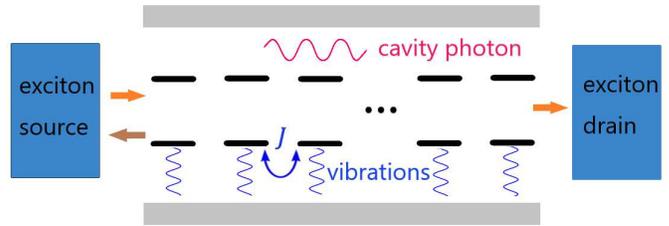}
\caption{Schematics of the source-molecular bridge-drain setup. A one-dimensional molecular aggregate located in a microcavity connects to an  exciton source and an exciton drain at its two ends. The electronic excitations of the monomers simultaneously interact with the cavity photon and the intramolecular vibrations.}
\label{Fig1}
\end{figure}
\begin{eqnarray}\label{SD-dis}
D_{\rm{r}}[\rho(t)]&=& \gamma_L \bar{n}_L\mathcal{L}_{a^\dag_1}(\rho)+ \gamma_L (\bar{n}_L+1)\mathcal{L}_{a_1}(\rho) + \gamma_R \mathcal{L}_{a_N}(\rho),\nonumber\\
\end{eqnarray}
where $\mathcal{L}_x(\rho)=x\rho x^\dag-\frac{1}{2}\{\rho,x^\dag x\}$ is the standard Lindblad superoperator. Here, $\gamma_L$ and $\gamma_R$ are the transfer rates related to the densities of states of the source and drain, respectively. $\bar{n}_{L}$ is the average exciton number in the exciton source.
The three terms on the right-hand side of Eq.~(\ref{SD-dis}) describe the exciton injection from the exciton source to monomer 1, the reflection of excitons back to the exciton source, and the exciton transmission from monomer $N$ to the drain, respectively. In addition, we also introduce the exciton decay ($\gamma_d$) due to the spontaneous emission, the exciton dephasing ($\gamma_p$) resulting from the coupling to the continuous phonon bath, and the decay of the cavity photon ($\kappa$) caused by the leakage through the mirrors:
\begin{eqnarray}\label{SD-dec}
D_{\rm{d}}[\rho(t)]&=&\sum^N_{n=1}[\gamma_d\mathcal{L}_{a_n}(\rho)+\gamma_p\mathcal{L}_{a^\dag_na_n}(\rho)]+\kappa \mathcal{L}_{c}(\rho).
\end{eqnarray}
\par Physically, the vibrational modes can also undergo relaxation at a rate of picoseconds~\cite{Plenio2015}, which is typically slower than that of the excitonic subsystem. The usual lifetime of the latter is in the timescale of several hundreds of femtoseconds~\cite{FJ2015}. We include this effect through Lindblad terms
\begin{eqnarray}\label{SD-v}
D_{\rm{v}}[\rho(t)]&=&\sum^N_{n=1}\gamma_v[(1+\bar{n}_v)\mathcal{L}_{b_n}(\rho)+\bar{n}_v\mathcal{L}_{b^\dag_n}(\rho)],
\end{eqnarray}
where the two terms on the right-hand side represent the damping and excitation processes of the vibrational modes due to local interaction with the phonon bath. Here, $\gamma_v$ is the relaxation rate and $\bar{n}_v$ is the mean bath occupation. For simplicity, we set $\bar{n}_v$ zero to include only the damping of the vibrations. We also restrict ourselves to the zero- and single-excitation subspace with $\sum_n a^\dag_n a_n+c^\dag c=0$ and $1$, so that we can write $a^\dag_n=|n\rangle_{\rm{e}}\langle \rm{vac}|$ and $c^\dag=|1\rangle_{\rm{c}}\langle \rm{vac}|$, where $|\rm{vac}\rangle$ is the common vacuum of the excitons and photons, and $|1\rangle_{\rm{c}}$ is the single-photon state~\cite{LPP}.
\par The dynamics of the whole system is thus governed by the master equation
\begin{eqnarray}\label{ME}
\frac{d\rho}{dt}=-i[H,\rho]+D_{\rm{r}}[\rho(t)]+D_{\rm{d}}[\rho(t)]+D_{\rm{v}}[\rho(t)].
\end{eqnarray}
Below we are interested in the steady-state exciton current in the molecular system.
\section{Exciton transport in the absence of vibrations}\label{SecIII}
To find out a proper definition of the exciton transfer efficiency, it is useful to first consider a system without vibrations. Such kind of models have been employed in several studies of exciton transport in organic materials strongly coupled to cavity modes~\cite{FJ2015,enhanced2015,Natm2015}. In the absence of the intramolecular vibrational modes, the model reduces to the Frenkel-Dicke model described by $H_{\rm{F-D}}=H_{\rm{e}}+H_{\rm{c}}+H_{\rm{e-c}}$, which generalizes the usual Dicke model by including dipole-dipole interactions between  nearest-neighbor monomers~\cite{Wu2018}.  Since we only consider the zero- and single-photon subspaces, $H_{\rm{F-D}}$ is further reduced to an interacting central spin model with spins-1/2~\cite{PRA2014}.
\par Let us first look at the dynamics for a simple molecular dimer bridge with $N=2$, for which the equations of motion for the matrix elements of $\rho$ can be written down explicitly (see Appendix \ref{AppA}). By investigating the population dynamics for the second monomer, Eq.~(\ref{rho22}), we see that one can naturally define the outgoing exciton current from the second monomer to the exciton drain as~\cite{Guan2013}
\begin{eqnarray}\label{Io_dimer}
I^{(\rm{dimer})}_{\rm{o}}&=&\gamma_R\rho_{22}=-\gamma_R\mathrm{Tr}\{a^\dag_2a_2 \mathcal{L}_{a_2}(\rho) \},
\end{eqnarray}
where the trace is taken over the whole system, i.e., the molecule-cavity system in the absence of vibrations.
\par For a general molecular chain, we analogously define the \emph{outgoing exciton current} from the $N$th monomer to the exciton drain as
\begin{eqnarray}\label{Io}
I_{\rm{o}}&=&-\gamma_R\mathrm{Tr}\{a^\dag_Na_N \mathcal{L}_{a_N}(\rho) \},
\end{eqnarray}
which is consistent with the current defined in Ref.~\cite{enhanced2015}. Note that the definition of the outgoing current given by Eq.~(\ref{Io}) is still valid in the presence of vibrations, with the understanding that the trace is taken over all the excitonic, photonic, and vibrational degrees of freedom. In a similar way, the \emph{utilizable input current} from the exciton source to the first monomer is defined as
\begin{eqnarray}\label{Ii}
I_{\rm{i}}&=&\gamma_L\mathrm{Tr}\{a^\dag_1a_1[(1+\bar{n}_L) \mathcal{L}_{a_1}(\rho)+\bar{n}_L \mathcal{L}_{a^\dag_1}(\rho)]\}.
\end{eqnarray}
We now label the vacuum state $|\mathrm{vac}\rangle$ of the exciton-cavity system as $|0\rangle$, and the $N+1$ exciton/cavity states as $|n\rangle=|n\rangle_{\mathrm{e}},~(n=1,2,\cdots,N)$ and $|N+1\rangle=|1\rangle_{\mathrm{c}}$. In the steady state, the difference between $I_{\rm{i}}$ and $I_{\rm{o}}$ gives the population decay rate
\begin{eqnarray}\label{Id}
I_{\rm{d}}&=&I_{\rm{i}}-I_{\rm{o}}=\gamma_d\sum^N_{n=1}\rho_{nn}+\kappa\rho_{N+1,N+1}\geq 0,
\end{eqnarray}
as can be seen from $\dot{\rho}_{00}=0$. Here, $\rho_{ij}=\langle i|\rho|j\rangle$ for $i,j=0,1,\cdots, N+1$.
In turn, we define the exciton transfer efficiency~\cite{Guan2013}
\begin{eqnarray}\label{eff}
\eta&=&\frac{I_{\rm{o}}}{I_{\rm{i}}}=\frac{I_{\rm{o}}}{I_{\rm{o}}+I_{\rm{d}}}\leq 1.
\end{eqnarray}
\begin{figure}
\includegraphics[width=.52\textwidth]{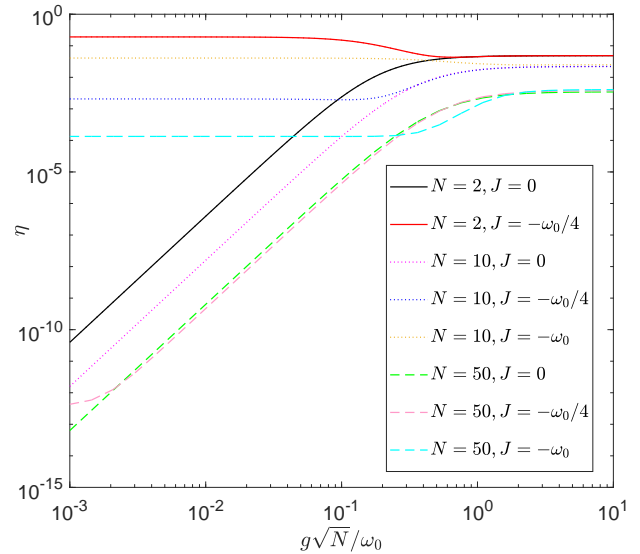}
\caption{Exciton transfer efficiency $\eta$ as a function of the collective exciton-cavity coupling $g\sqrt{N}/\omega_0$, in the absence of the exciton-vibration coupling. Results for molecular chains with $N=2, 10$, and $50$ monomers are presented. Other parameters: $\varepsilon_n=\omega_{\mathrm{c}}=2$ eV, $\omega_0=0.17$ eV, $\lambda=0$, $\gamma^{-1}_d=600 ~\mathrm{fs}$, $\gamma^{-1}_p=25 ~\mathrm{fs}$, $\kappa^{-1}=50 ~\mathrm{fs}$, $\gamma^{-1}_L=\gamma^{-1}_R=1 ~\mathrm{ps}$, and $\bar{n}_L=1$.}
\label{Fig2}
\end{figure}
\par Figure \ref{Fig2} shows the exciton transfer efficiency $\eta$ through a molecular chain with different sizes and for different hopping integral $J$, as a function of the dimensionless collective exciton-cavity coupling $g\sqrt{N}/\omega_0$. In the numerical simulation, we choose the system parameters approximately corresponding to the J aggregates at room temperature~\cite{FJ2015}: $\varepsilon_n=2$ eV, $\omega_0=0.17$ eV, $\gamma^{-1}_d=600 ~\mathrm{fs}$, $\gamma^{-1}_p=25 ~\mathrm{fs}$, and $\bar{n}_L=1$. The cavity is assumed to be resonant with the on-site excitonic transition with frequency $\omega_{\mathrm{c}}=2$ eV and lifetime $\kappa^{-1}=50 ~\mathrm{fs}$. The pumping rate is set to be $\gamma^{-1}_L=\gamma^{-1}_R=1 ~\mathrm{ps}$ to guarantee the validity of the single-excitation approximation~\cite{FJ2015}.
\par It can be seen that the exciton transfer efficiency $\eta$ is mainly affected by the hopping integral $J$ before the onset of strong cavity coupling. 
In the strong-coupling regime, an extraordinary increase of the transfer efficiency is observed. As $|J|$ exceeds some turnover value, the strong-coupling enhancement disappears since direct transport through exciton hopping dominates. The turnover takes place at larger values of $|J|$ for increasing numbers of monomers $N$. As pointed out in Ref.~\cite{FJ2015}, these observations can be explained through two almost independent transfer channels, i.e., the hopping-dominated direct transfer in the weak cavity coupling limit, and a polaritonic transport through polariton modes created in the strong coupling regime.
\par In the absence of vibrations, we follow Ref.~\cite{FJ2015} to define the onset of strong exciton-cavity coupling occurring at
\begin{eqnarray}
g_c\sqrt{N}=|\gamma_d+\gamma_p-\kappa|/4.
\end{eqnarray}
For the parameters used in the present work, we have $g_c\sqrt{N}/\omega_0\approx 0.132$. Below we follow this to define the parameter range $g\sqrt{N}/\omega_0<0.132$ ($>0.132$) as the weak (strong) exciton-cavity coupling region, even in the presence of vibrations.
\section{Exciton transport in the presence of vibrations}\label{SecIV}
Given the input and output exciton current defined above, we now turn to the study of vibrational effects on the exciton transport. We first consider exciton transport through a molecular dimer, which has been employed to investigate several phenomena in light-harvesting systems~\cite{Guan2013,NC2014,Plenio2015}. We then focus on molecular chains with $N\geq 4$ monomers to investigate the length dependence of the exciton transfer process in the presence of vibrational modes.

\par It is known that the real-time dynamics of the Holstein model with infinitely many bosonic degrees of freedom is notoriously difficult to treat, even though a variety of numerical or analytical methods have been proposed to deal with its zero-temperature unitary dynamics~\cite{Zhang,Vidm,chin}. To this end, in the following numerical simulations we choose to truncate the vibrational space by keeping only limited total numbers of vibrations $\hat{M}_{\max}=\sum^N_{i=1}b^\dag_i b_i$ in the molecular chain. For example, for $M_{\max}=2$ the truncated vibrational space is spanned by the vibrational vacuum $|0,\cdots,0\rangle_{\mathrm{v}}$, the $N$ single-vibration states $|1,0,\cdots,0\rangle_{\mathrm{v}},~|0,1,\cdots,0\rangle_{\mathrm{v}},\cdots$, as well as the $N(N+1)/2$ two-vibration states $|2,0,\cdots,0\rangle_{\mathrm{v}},~|0,2,\cdots,0\rangle_{\mathrm{v}},\cdots$ and $|1,1,0,\cdots,\rangle_{\mathrm{v}}, |1,0,1,\cdots,\rangle_{\mathrm{v}},\cdots$. The total dimension of this truncated subspace is thus $(N+1)(N+2)/2$.

\subsection{Molecular dimer}

In this subsection, we study in particular a molecular dimer as a bridge for exciton transport. In the absence of the cavity, the two-site Holstein model describing the molecular dimer can be mapped to the single-mode Rabi model by introducing the centre of mass mode and relative displacement mode of the vibrations~\cite{NC2014}.
\begin{figure}
(a)
\includegraphics[width=.50\textwidth]{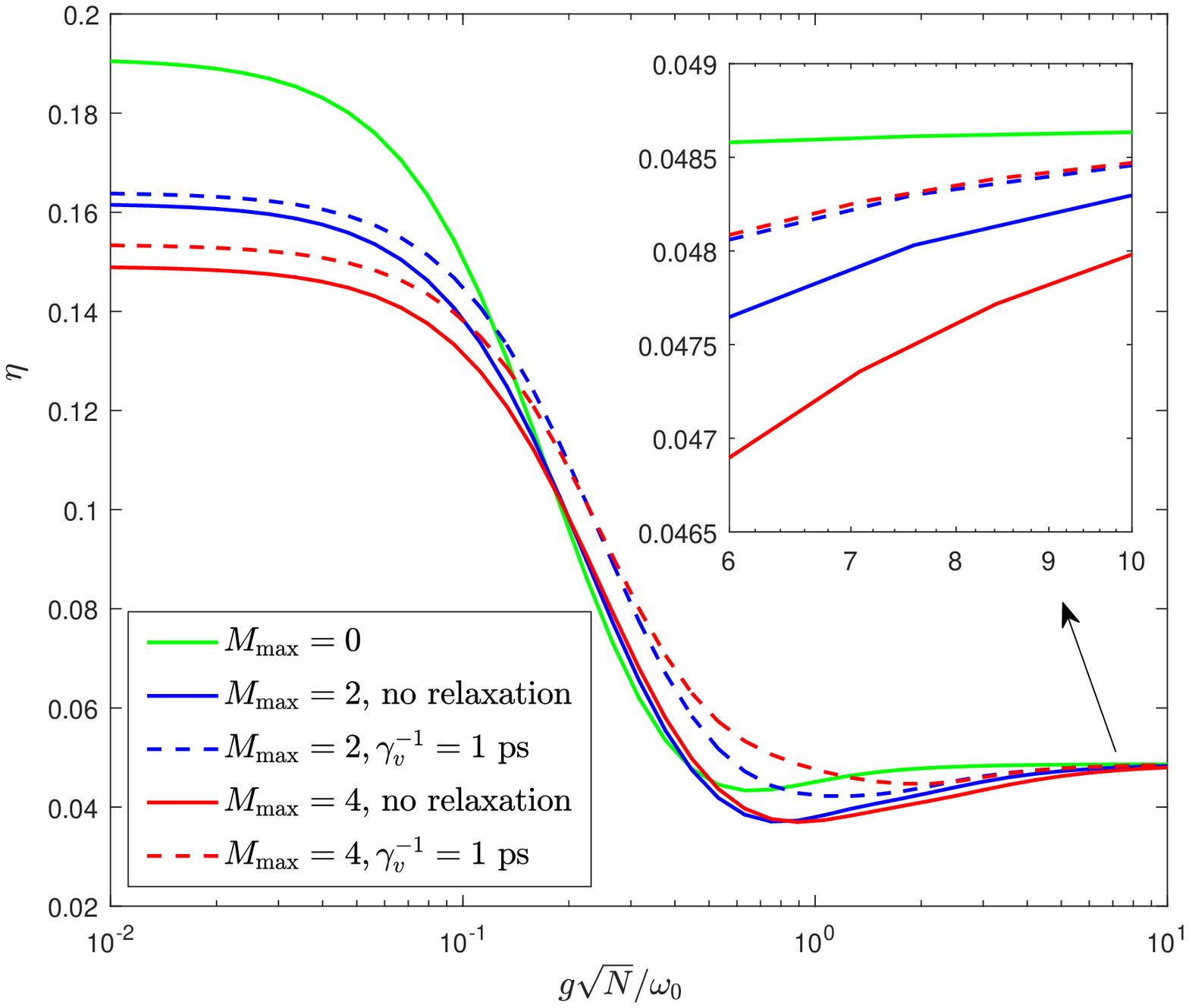}
(b)
\includegraphics[width=.50\textwidth]{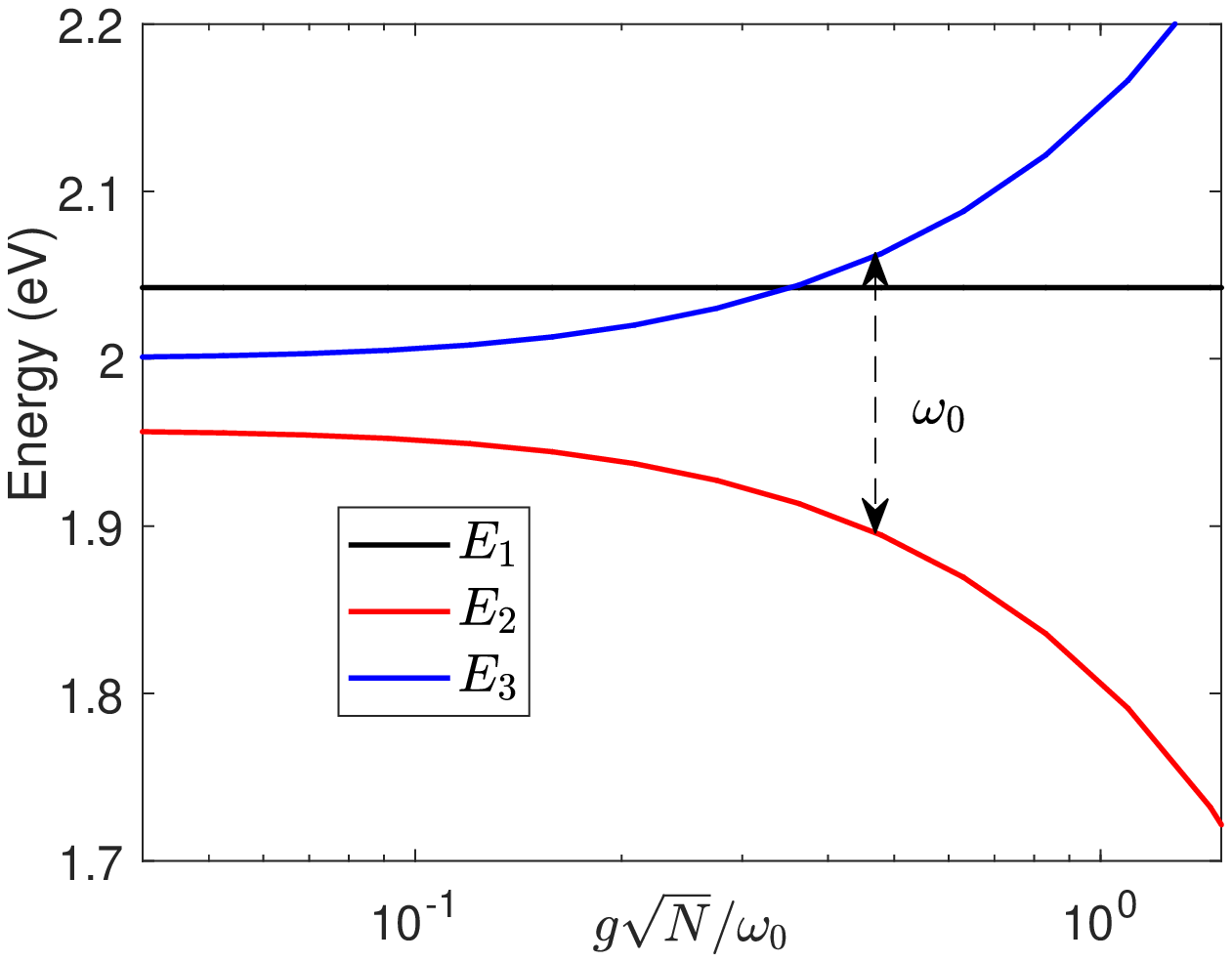}
\caption{(a) Exciton transfer efficiency $\eta$ as a function of $g\sqrt{N}/\omega_0$ for a molecular dimer in the presence of different total numbers of vibrational quanta. The inset shows magnification in the strong coupling regime with $g\sqrt{N}/\omega_0\geq6$. Other parameters: $\varepsilon_n=\omega_{\mathrm{c}}=2$ eV, $\omega_0=0.17$ eV, $J=-\omega_0/4$, $\lambda^2=1$, $\gamma^{-1}_d=600 ~\mathrm{fs}$, $\gamma^{-1}_p=25 ~\mathrm{fs}$, $\kappa^{-1}=50 ~\mathrm{fs}$, $\gamma^{-1}_L=\gamma^{-1}_R=1 ~\mathrm{ps}$, and $\bar{n}_L=1$. (b) Eigenenergies of the bare exciton-cavity system as functions of $g\sqrt{N}/\omega_0$. The energy gap between the upper polariton (blue) and the lower polariton (red) matches the vibrational frequency $\omega_0$ at $g\sqrt{2}/\omega_0\approx 0.484$.}
\label{Fig3}
\end{figure}
However, in the presence of the cavity, the excitonic and photonic degrees of freedom are mixed and hence make this decoupling impossible.  The truncated vibrational space has dimension $D_v=(M_{\max}+1)(M_{\max}+2)/2$.

\par Figure \ref{Fig3}(a) shows the evolution of the profile of $\eta$ when the truncated total number of vibrations is increased, up to $M_{\max}=4$ vibrational quanta. The Huang-Rhys factor is chosen as $\lambda^2=1$, indicating that the system lies in the strong exciton-vibration coupling regime. In the absence of the vibrational relaxation ($\gamma_v=0$ eV, solid curves), it can be seen that except for the minor enhancement of the efficiency for exciton-cavity couplings around $g\sqrt{N}/\omega_0\sim 0.5$, the introduction of vibrational quanta will suppress the exciton transfer in both the weak and strong coupling limits. Nevertheless, we observe an overall enhancement of $\eta$ by including finite relaxation of vibrations with rate $\gamma_v^{-1}=1~\mathrm{ps}$ ($\gamma_v=4.14$ meV, dashed curves), especially in the intermediate to strong exciton-cavity coupling regime. The vibrational suppression and relaxation-enhancement even hold in the ultra-strong coupling limit, where the vibrational effects tend to be washed out and the transfer efficiency approaches the result for the bare exciton-cavity system [inset of Fig.~\ref{Fig3}(a)].
\begin{figure}
\includegraphics[width=.55\textwidth]{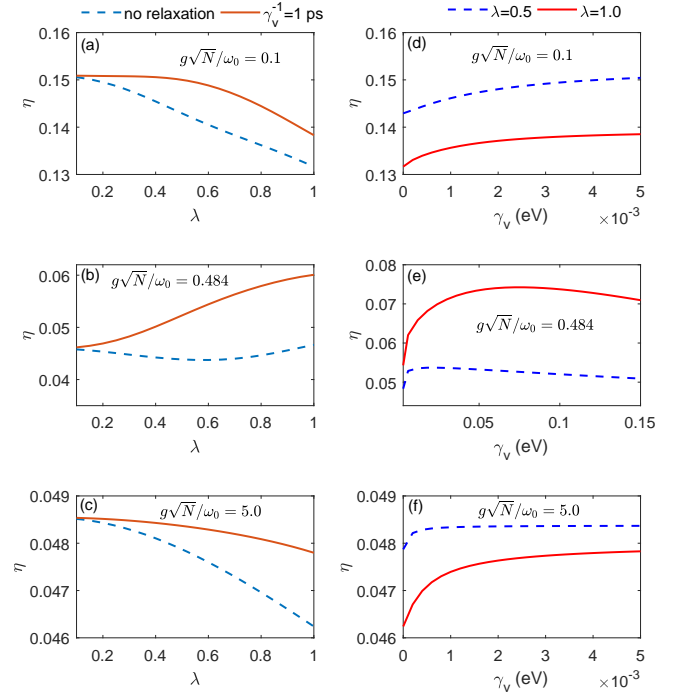}
\caption{Evolution of the exciton transfer efficiency $\eta$ for a molecular dimer with the dimensionless exciton-vibration coupling $\lambda$ (left column) and the relaxation rate $\gamma_v$ of the vibrational modes (right column), for different exciton-cavity coupling strengths $g\sqrt{N}/\omega_0= 0.1$ (first row), $0.484$ (middle row), and $5.0$ (last row). Other parameters: $M_{\max}=4$, $\varepsilon_n=\omega_{\mathrm{c}}=2$ eV, $\omega_0=0.17$ eV, $J=-\omega_0/4$, $\gamma^{-1}_d=600 $ fs, $\gamma^{-1}_p=25$ fs, $\kappa^{-1}=50 $ fs, $\gamma^{-1}_L=\gamma^{-1}_R=1$ ps, and $\bar{n}_L=1$.}
\label{Fig4}
\end{figure}
\par To understand these observations, let us study the bare exciton-cavity system in the absence of vibrations. In this case, the Frenkel-Dicke Hamiltonian $H_{\mathrm{F-D}}(N=2)$ for a homogeneous dimer coupled to a resonant cavity (with $\varepsilon_1=\varepsilon_2=\omega_{\mathrm{c}}$) can be diagonalized analytically in the basis $\{|1\rangle_{\mathrm{e}},|2\rangle_{\mathrm{e}},|1\rangle_{\mathrm{c}}\}$, yielding the following three eigenstates
\begin{eqnarray}
|\psi_1\rangle&=&\frac{1}{\sqrt{2}}\left(
                                     \begin{array}{c}
                                       1 \\
                                       -1 \\
                                       0 \\
                                     \end{array}
                                   \right)
,\nonumber\\
|\psi_2\rangle&=&\frac{1}{2\sqrt{J^2+8g^2-J\sqrt{J^2+8g^2}}}\left(
                                     \begin{array}{c}
                                        J-\sqrt{J^2+8g^2} \\
                                        J-\sqrt{J^2+8g^2} \\
                                       4g\\
                                     \end{array}
                                   \right),\nonumber\\
       |\psi_3\rangle&=&\frac{1}{2\sqrt{J^2+8g^2+J\sqrt{J^2+8g^2}}}\left(
                                     \begin{array}{c}
                                        J+\sqrt{J^2+8g^2} \\
                                        J+\sqrt{J^2+8g^2} \\
                                       4g\\
                                     \end{array}
                                   \right).\nonumber
\end{eqnarray}
The corresponding eigenenergies are
\begin{eqnarray}
E_1&=&\omega_c-J,\nonumber\\
E_2&=&\omega_c+\frac{J}{2}-\frac{1}{2}\sqrt{J^2+8g^2},\nonumber\\
E_3&=&\omega_c+\frac{J}{2}+\frac{1}{2}\sqrt{J^2+8g^2}.
\end{eqnarray}
It can be seen that $|\psi_1\rangle$ is a dark state independent of the cavity mode, while only $|\psi_2\rangle$ and $|\psi_3\rangle$ have the photonic component, with their energy spacing $\Delta E_{32}=E_3-E_2=\sqrt{J^2+8g^2}$. The latter two states, respectively known as the lower and upper polariton states, arise for both dimers and large ensembles when $g\sqrt{N}/\omega_0$ is sufficiently large to overcome the lineshape broadening~\cite{Torma} and observe their so-called Rabi splitting ($\Delta E_{32}$ here).
\par The effects of vibrational modes are expected to be most important when the vacuum Rabi splitting matches the frequency of a single vibrational quanta, i.e., $\Delta E_{32}\approx\omega_0$.  This results in $g\sqrt{2}/\omega_0\approx 0.484$, which just lies in the vibrational enhancement region [Fig.~\ref{Fig3}(b)]. In this case, if we consider the population dynamics starting with certain prepared initial states, e.g., the upper polariton state $|\psi_3\rangle$ (such kind of polariton dynamics has been studied in Ref.~\cite{chin}), then the presence of vibrations will open a new decay pathway to the lower polariton $|\psi_2\rangle$ through emitting a single vibration~\cite{NC2014,Plenio2015,PRX,chin}. However, coherent exchange can also transfer excitation back from $|\psi_2\rangle$ to $|\psi_3\rangle$. Thanks to the dissipative processes on the vibrational mode, finite relaxation of the mode will suppress disadvantageous back-transfer, and hence makes the excitation transfer to lower-lying states directional.

\par To better understand the vibrational effects on the stationary exciton transfer through a molecular dimer, we plot in Fig.~\ref{Fig4} the dependence of the efficiency $\eta$ on the dimensionless exciton-vibration coupling strength $\lambda$ and on the relaxation rate $\gamma_v$. Results for $g\sqrt{N}/\omega_0=0.1,~0.484,$ and $5.0$ are shown as representatives of the weak, strong, and ultrastrong coupling regime, respectively. We used $M_{\max}=4$ for the truncated vibrational space. The left column of Fig.~\ref{Fig4} shows the evolution of $\eta$ as the exciton-vibration coupling increases. It can be seen that increasing $\lambda$ will generally suppress the transfer efficiency in the weak and ultrastrong coupling limits. However, we observe vibrational enhancement for the matching coupling $g\sqrt{N}/\omega_0=0.484$ [for finite mode relaxation rate $\gamma_v=(1~\mathrm{ps})^{-1}$, Fig.~\ref{Fig4}(b)]. The right column of Fig.~\ref{Fig4} shows the corresponding transfer efficiency when the relaxation rate $\gamma_v$ is increased. The environment induced relaxation of the vibrational modes is always found to facilitate the exciton transfer process for $\gamma_v$ up to $5~\mathrm{meV}$. A closer look at Fig.~\ref{Fig4}(e) reveals that for $g\sqrt{N}/\omega_0=0.484$ with $\lambda=1.0$, there actually exists an optimal relaxation rate $\gamma_v\approx 70~\mathrm{meV}$ (about $60~\mathrm{fs}^{-1}$) beyond which the transfer efficiency gradually decreases. Note that this relaxation rate is comparable to the fast decay process of the exciton-cavity system, i.e., the photon decay rate $\kappa^{-1}=50~\mathrm{fs}$.
\par
\begin{figure}
\includegraphics[width=.50\textwidth]{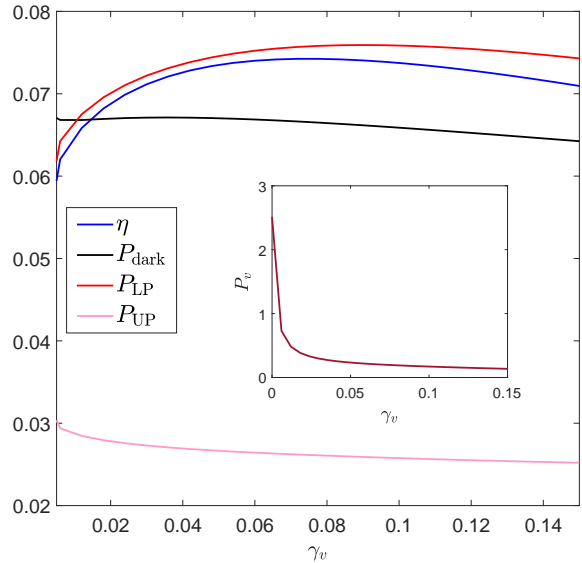}
\caption{Steady-state occupations of the dark state, the lower polariton state, and the upper polariton state for a molecular dimer as functions of the relaxation rate $\gamma_v$ on the vibrations. The inset shows the corresponding numbers of vibrations in the steady state. The exciton-cavity coupling is chosen as $g\sqrt{N}/\omega_0=0.484$ to see the vibrational enhancement of the transport. Other parameters: $\varepsilon_n=\omega_{\mathrm{c}}=2$ eV, $\omega_0=0.17$ eV, $J=-\omega_0/4$, $\lambda^2=1$, $\gamma^{-1}_d=600 $ fs, $\gamma^{-1}_p=25$ fs, $\kappa^{-1}=50 $ fs, $\gamma^{-1}_L=\gamma^{-1}_R=1$ ps, and $\bar{n}_L=1$. }
\label{occupation}
\end{figure}

 Since the unavoidable decay of the exciton-cavity system always tends to send the excitation to the vacuum state, the presence of vibrations is expected to compete against these loss mechanisms if the transport is vibrationally enhanced. To this end, we plot in Fig.~\ref{occupation} the stationary occupations of the dark state $|\psi_1\rangle$, the lower polariton $|\psi_2\rangle$, and the upper polariton $|\psi_3\rangle$ in the steady state as functions of the relaxation rate $\gamma_v$. These quantities are respectively given by $P_{\rm{dark}}=\mathrm{Tr}(\rho_{ss}|\psi_1\rangle\langle \psi_1| )$, $P_{\rm{LP}}=\mathrm{Tr}(\rho_{ss}|\psi_2\rangle\langle \psi_2| )$, and $P_{\rm{UP}}=\mathrm{Tr}(\rho_{ss}|\psi_3\rangle\langle \psi_3| )$, where $\rho_{ss}$ is the steady-state density matrix. As $\gamma_v$ increases, though the upper polariton population decreases monotonically, the occupations of the lowest two states, the lower polariton and the dark state, are found to display similar nonmonotonic behavior with the transfer efficiency $\eta$. From the point of view of detailed balance, lower-lying states tend to acquire higher occupations in the steady state, and hence contribute more to the transport process. Due to the excitation decay at a fixed rate, the excitation energy transfer must occur at a shorter time scale than the fast decay processes. As a linear combination of the photonic and excitonic states, the lower polariton experiences the spontaneous emission and the cavity decay simultaneously, which explains the stronger correlation between its occupation and the transfer efficiency.

\par  When $\gamma_v$ increases from zero to the turning point $\gamma_v\approx 70~\mathrm{meV}$, the progressively enhanced damping of the vibrations helps to redistribute the occupations of the three states so as to facilitate the transport by competing with the decay. However, once the relaxation rate is increased further and go beyond the optimal point, the vibrational enhancement will be weakened due to the loss of vibrations. The inset of Fig.~\ref{occupation} shows that the total numbers of vibrations in the steady state, $P_v=\mathrm{Tr}[\rho_{ss}(b^\dag_1b_1+b^\dag_2 b_2)]$, decrease monotonically as we increase $\gamma_v$. We see that the vibrational enhancement does not show positive correlation with the total numbers of vibrations. We attribute the vibrational suppression of the transfer efficiency in the weak and ultrastrong coupling regimes to the mismatch between the Rabi splitting and the vibration's frequency.

We also note that the introduction of vibrational modes does not influence the phenomenon that ultrastrong cavity coupling suppresses the exciton transfer through short chains such as a molecular dimer (Fig.~\ref{Fig3}). This is because in this regime the vacuum Rabi splitting is much larger than the frequency of the vibrations, so that vibrational effects only plays a minor role in the exciton transport.
\subsection{Molecular chains}
\begin{figure}
\includegraphics[width=.50\textwidth]{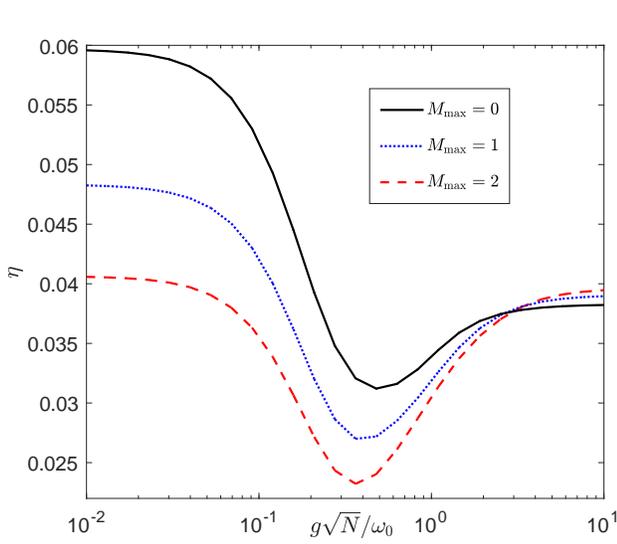}
\caption{Evolution of the exciton transfer efficiency $\eta$ with increasing numbers of vibrational quanta $M_{\max}$ for a molecular chain with $N=4$ monomers.
Other parameters: $\varepsilon_n=\omega_{\mathrm{c}}=2$ eV, $\omega_0=0.17$ eV, $J=-\omega_0/4$, $\lambda^2=1$, $\gamma_v=0$, $\gamma^{-1}_d=600 $ fs, $\gamma^{-1}_p=25$ fs, $\kappa^{-1}=50 $ fs, $\gamma^{-1}_L=\gamma^{-1}_R=1$ ps, and $\bar{n}_L=1$.}
\label{Fig6}
\end{figure}
For longer molecular chains with $N\geq 4$ monomers, the dimension of the vibrational space increases rapidly with $N$. For simplicity, we therefore only include at most $M_{\max}=2$ total vibrational quanta in the following simulations of exciton transport through molecular chains. Figure~\ref{Fig6} shows the exciton transfer efficiency for a molecular chain with $N=4$ monomers. Results for $M_{\max}=0,~1$, and $2$ demonstrate the influence of increasing the number of vibrations. The relaxation rate $\gamma_v$ is set zero.
\begin{figure}
(a)
\includegraphics[width=.51\textwidth]{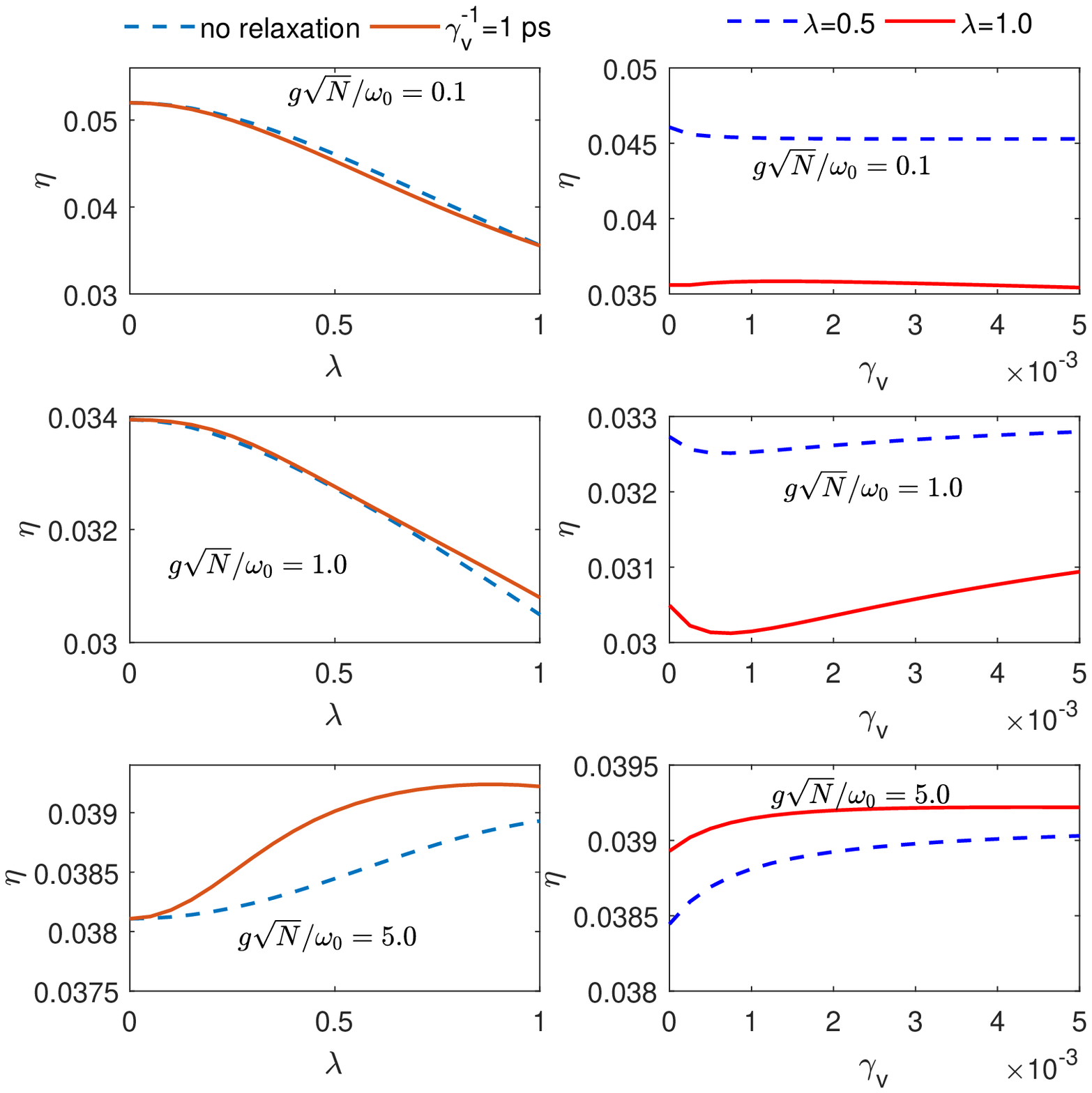}
(b)
\includegraphics[width=.52\textwidth]{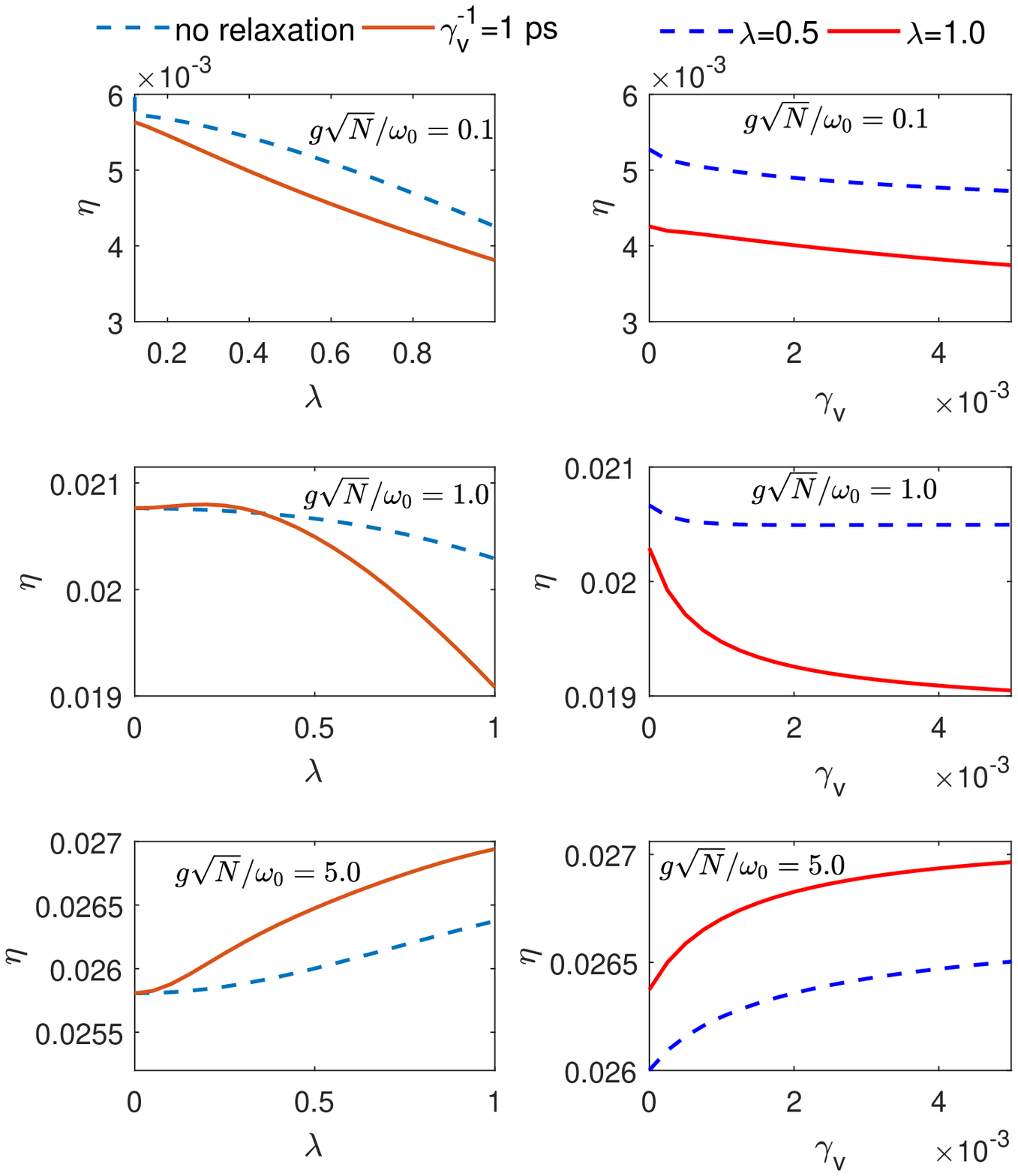}
\caption{The same as in Fig.~\ref{Fig4}, but for a molecular chains with (a) $N=4$ monomers, $M_{\max}=2$ vibrational quanta, and (b) $N=8$ monomers, $M_{\max}=1$ vibrational quanta. Other parameters: $\varepsilon_n=\omega_{\mathrm{c}}=2$ eV, $\omega_0=0.17$ eV, $J=-\omega_0/4$, $\gamma^{-1}_d=600 ~\mathrm{fs}$, $\gamma^{-1}_p=25 ~\mathrm{fs}$, $\kappa^{-1}=50 ~\mathrm{fs}$, $\gamma^{-1}_L=\gamma^{-1}_R=1 ~\mathrm{ps}$,  and $\bar{n}_L=1$.}
\label{Fig7}
\end{figure}
From the weak- to strong-coupling regime, increasing the introduced total number of vibrations results in significant suppression of the transfer efficiency, which even holds up to relatively strong exciton-cavity coupling with $g\sqrt{N}/\omega_0\sim1$. However, as the exciton-cavity coupling increases further, a crossover to a vibrational enhancement region takes place. In this ultrastrong-coupling regime, the transfer efficiency increases with increasing $M_{\max}$, though the magnitude of increment is much smaller compared to the corresponding decrement in the weak- and strong-coupling regimes. In fact, the efficiency is converging.
\begin{figure}
\includegraphics[width=.51\textwidth]{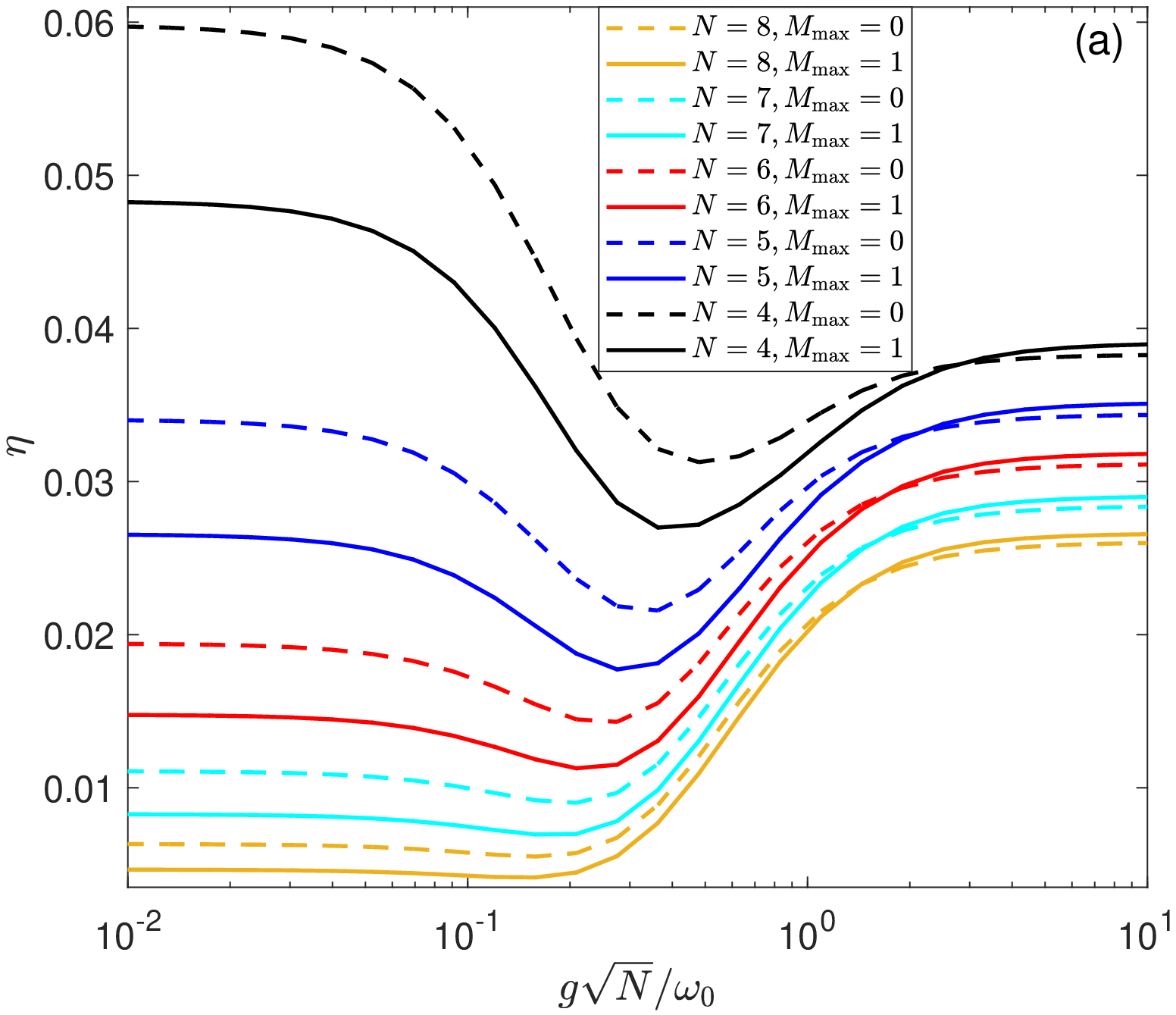}
\includegraphics[width=.51\textwidth]{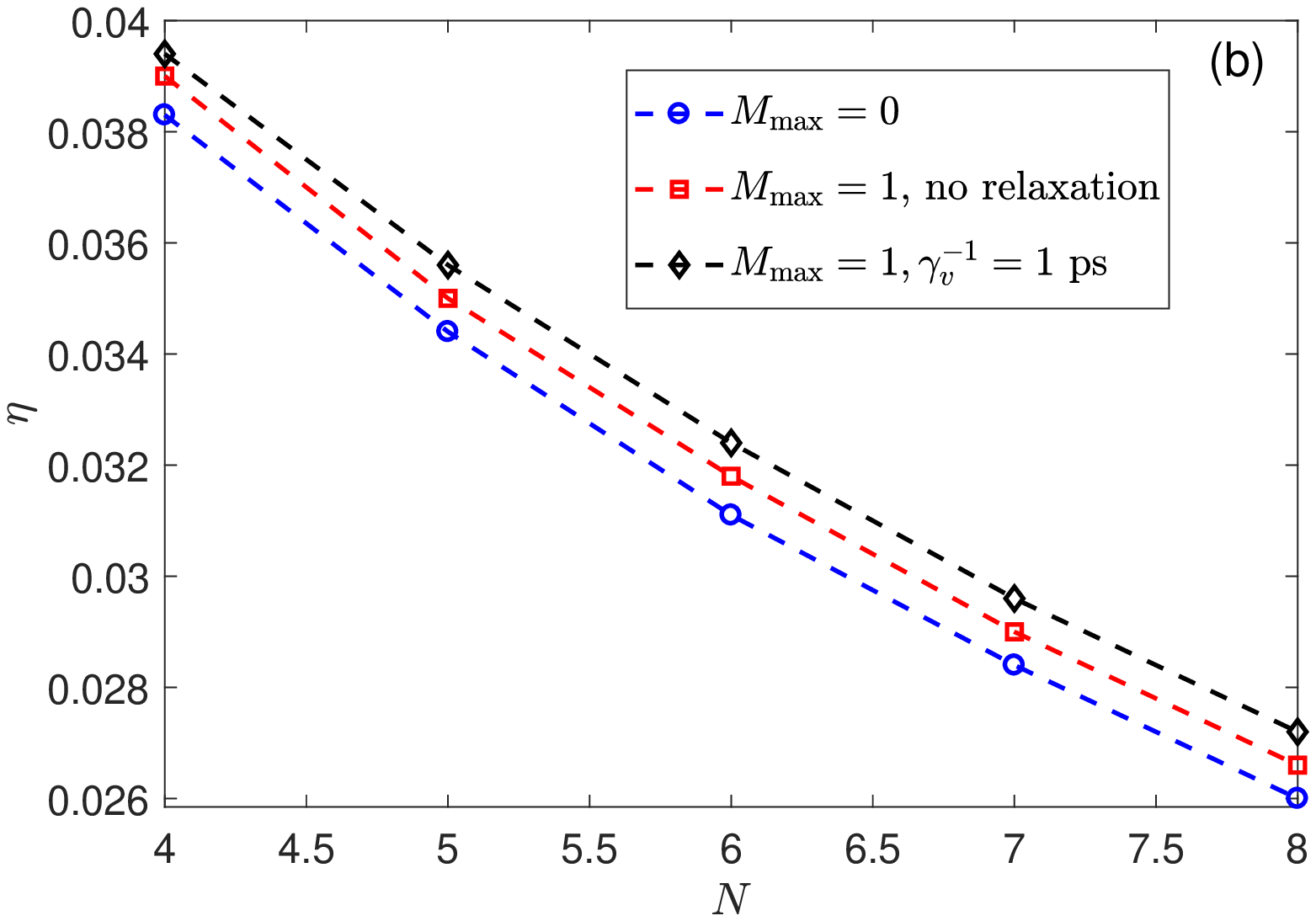}
\caption{(a) Exciton transfer efficiency $\eta$ for molecular chains with different sizes. The dashed (solid) curves show the results with no (one) vibration included. Here, $\gamma_v=0$ eV is used. (b) Exciton transfer efficiency $\eta$ as a function of the number of monomers $N$ at strong exciton-cavity coupling $g\sqrt{N}/\omega_0=5.0$.
Other parameters: $\varepsilon_n=\omega_{\mathrm{c}}=2$ eV, $\omega_0=0.17$ eV, $J=-\omega_0/4$, $\lambda^2=1$, $\gamma^{-1}_d=600 ~\mathrm{fs}$, $\gamma^{-1}_p=25 ~\mathrm{fs}$, $\kappa^{-1}=50 ~\mathrm{fs}$, $\gamma^{-1}_L=\gamma^{-1}_R=1 ~\mathrm{ps}$,  and $\bar{n}_L=1$.}
\label{Fig8}
\end{figure}

\par We plot in Fig.~\ref{Fig7}(a) the dependence of $\eta$ on $\lambda$ and $\gamma_v$ for $N=4$ and $M_{\max}=2$. It can be seen that increasing the exciton-vibrational coupling $\lambda$ suppresses the exciton transfer from weak to moderately strong coupling regimes [first two rows of Fig.~\ref{Fig7}(a)]. In contrast, we observe enhanced transfer efficiency by the exciton-vibration coupling in the ultrastrong exciton-cavity coupling limit [last row of Fig.~\ref{Fig7}(a)], where relaxation of the vibrations also helps to increase the transfer efficiency. For molecular chains with $N\geq5$ monomers, our simulations are limited to the single-vibration subspace with $M_{\max}=1$. Figure~\ref{Fig7}(b) shows the transfer efficiency $\eta$ as functions of $\lambda$ and $\gamma_v$ for a molecular chain with $N=8$ monomers, where $M_{\max}=1$ is used. As found for $N=4$, for ultrastrong cavity couplings, increasing the exciton-vibration coupling and the relaxation rate of the vibrational modes both facilitate the exciton transfer [last row of Fig.~\ref{Fig7}(b)].

\par Figure~\ref{Fig8}(a) shows the evolution of the profile of $\eta$ with the number of monomers increased. The dashed and solid curves correspond to cases without vibrations and with single-vibration states included, respectively. We first note that in the weak- to strong-coupling regime the discrepancy between the dashed curves and the corresponding solid ones becomes smaller and smaller with increasing $N$, indicating that the vibrational correction to $\eta$ becomes less significant for long molecular chains in this regime. However, the increment of $\eta$ caused by the vibrational modes in the ultrastrong coupling limit is almost independent of the system size, as can be seen from Fig.~\ref{Fig8}(b).

\par
\begin{figure}
(a)
\includegraphics[width=.52\textwidth]{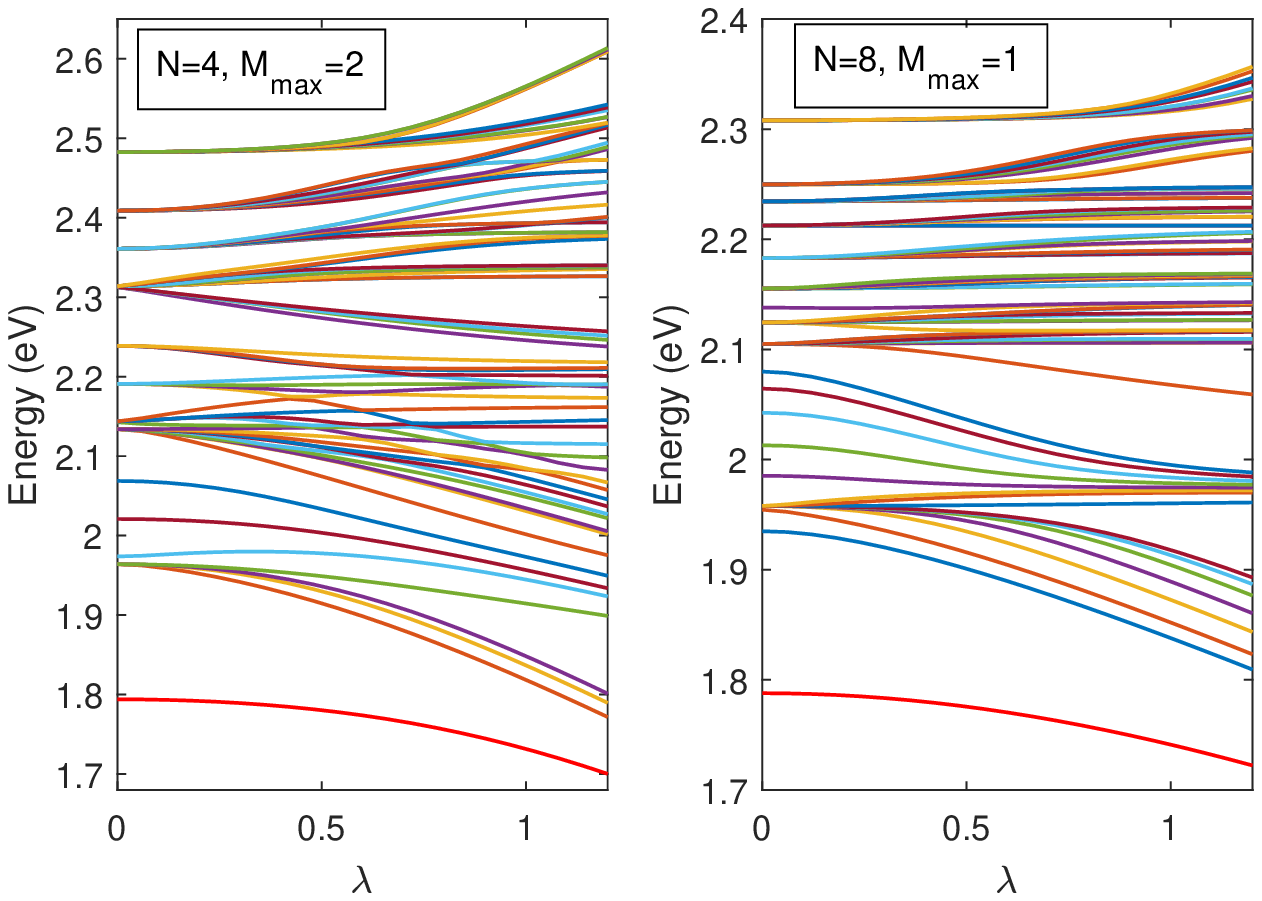}
(b)
\includegraphics[width=.54\textwidth]{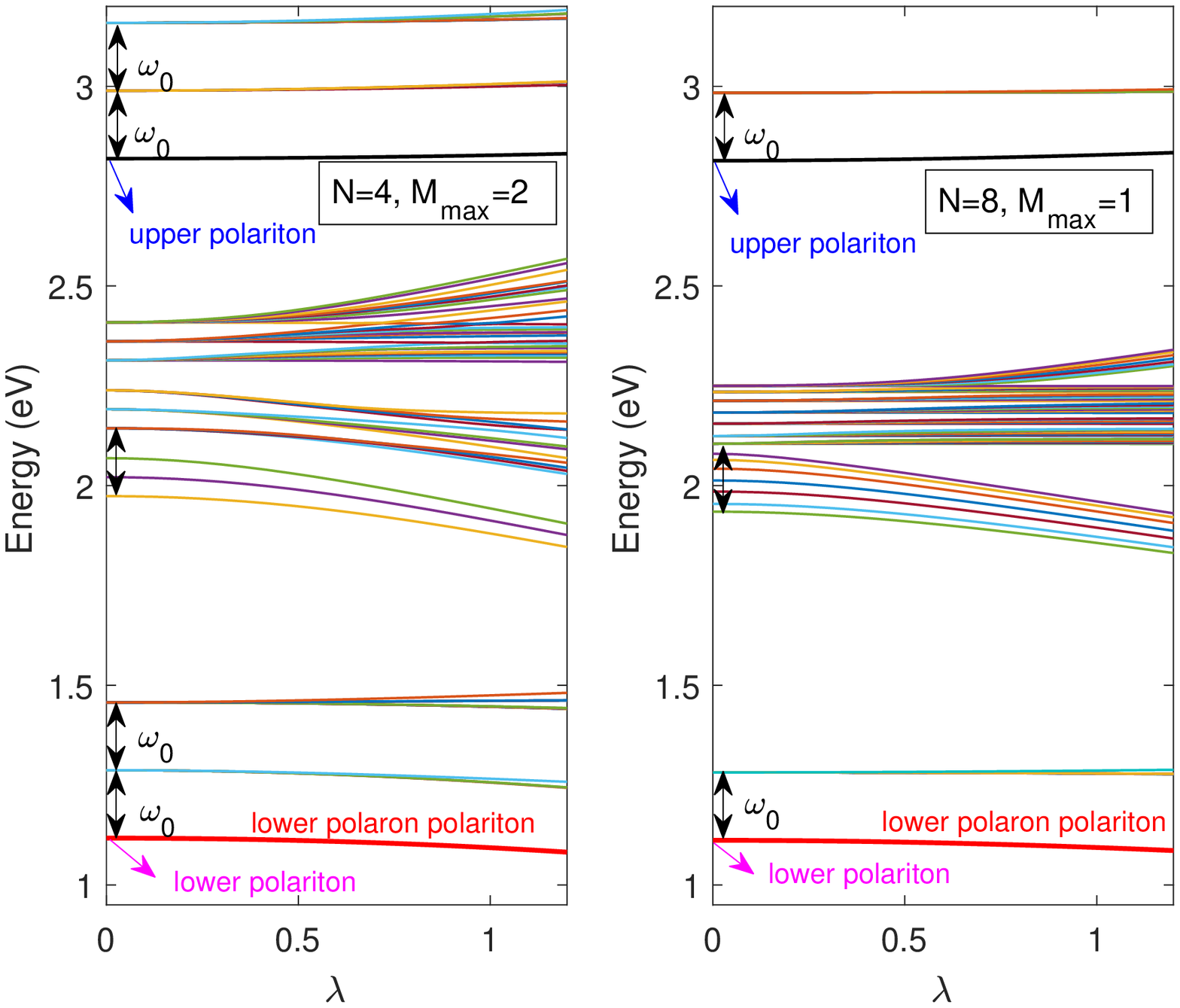}
\caption{Energy spectrum of the exciton-photon-vibration hybrid system for $N=4$ ($M_{\max}=2$ is used, left panels) and $N=8$ ($M_{\max}=1$ is used, right panels) at exciton-cavity coupling strength (a) $g\sqrt{N}/\omega_0=1.0$, and (b) $g\sqrt{N}/\omega_0=5.0$. The energy levels corresponding to the upper polariton and lower polariton, which lie on the $\lambda=0$ line, are indicated by blue and pink arrows, respectively. Other parameters: $\varepsilon_n=\omega_{\mathrm{c}}=2$ eV, $\omega_0=0.17$ eV, and $J=-\omega_0/4$.}
\label{Fig9}
\end{figure}

 To qualitatively understand the phenomenon of vibration-assisted exciton transport in the ultrastrong cavity coupling regime, we present in Fig.~\ref{Fig9} the full spectrum of the whole system as a function of $\lambda$ for both $g\sqrt{N}/\omega_0=1.0$ [Fig.~\ref{Fig9}(a)] and $g\sqrt{N}/\omega_0=5.0$ [Fig.~\ref{Fig9}(b)]. In the weak exciton-vibration coupling limit $\lambda\to0$, the vibrations are decoupled from the exciton-cavity system. Thus, the excitonic dynamics is controlled by the $N+1$ bare exciton-photon hybrid states. In the strong cavity coupling limit, these states are divided into a quasi-continuous band containing $N-1$ exciton-dominated states with bandwidth $\sim |4J|=\omega_0$, as well as  the so-called lower and upper polaritons [indicated by arrows in Fig.~\ref{Fig9} (b)]. It should be noted that in the $\lambda\to 0$ limit the bandwidth of the excitonic band is almost independent of the exciton-cavity coupling~\cite{LPP}. The emerging band structure of dark states for longer chains is in contrast to the case of the molecular dimer, where only a single dark state exists.

In the presence of the exciton-vibration coupling, the vibrational modes get mixed with the exciton-photon system, resulting in complicated spectrum structure that involves all the three types of degrees of freedom. For exciton-cavity coupling with strength $g\sqrt{N}/\omega_0=1.0$, as $\lambda$ increases, the exciton-photon states develop into several irregular broad quasi-continuous bands with band gaps smaller than $\omega_0$ [Fig.~\ref{Fig9}(a)]. However, in the ultrastrong coupling regime with $g\sqrt{N}/\omega_0=5.0$, the presence of the exciton-vibration coupling leads to new vibrational bands on top of the original bare states. In particular, the lower polariton state [pink arrows in Fig.~\ref{Fig9}(b)] develops into the so-called lower polaron polariton in the strong exciton-cavity and strong exciton-vibration coupling regime~\cite{LPP}, whose energy is pulled down as $\lambda$ increases (the lower red curves). The lower polaron polariton state, which was firstly named in Ref.~\cite{LPP}, is defined as the ground state of the hybrid Hamiltonian $H_{\rm{Hol}}+H_{\rm{c}}+H_{\rm{e-c}}$ in the single excitonic and photonic excitation subspace, and it involves all the excitonic, photonic, and vibrational degrees of freedom. The states developed from the bare upper polariton [blue arrows in Fig.~\ref{Fig9}(b)] tend to be lifted up with increasing $\lambda$. In this ultrastrong coupling regime, both the lower polaron polariton states and the states emerging from the upper polariton are well separated from the middle bands. In addition, the bandwidth of the subbands evolving from the remaining $N-1$ exciton-dominated states tends to be narrowed down as $\lambda$ increases, and the overall profile is pulled downward. However, the newly generated vibrational subbands associated with these dark states are broadened by increasing exciton-vibrational coupling, and more fine structures of the spectrum emerge in the strong exciton-vibration coupling regime.
\par Compared with the dimer for which the behavior of occupations of lower-lying excited states could serve as an indicator of the transfer efficiency, it is less straightforward to describe the transport mechanism through molecular chains due to the complex interplay of the exciton-vibration coupling. The master equation describing the exciton transport through a molecular chain involves not only the populations of the eigenstates (corresponding to diagonal elements of the density matrix), but also mutual interference between them. Since the dimension of the Hilbert space increases rapidly for longer molecular chains, the off-diagonal elements of the steady-state density matrix might play an important role in determining transport properties for longer molecular chains.

\par The band gaps between adjacent subbands are roughly $\omega_0$, so that transition to the upper subband through absorbing a vibrational quantum becomes possible. When the exciton-vibration coupling is increased, the vibrationally dressed dark states in the upper-lying subbands come closer together and begin to overlap. Once the relaxation $\gamma_v$ of vibrations is large enough to suppress the population back-transfer to lower subbands, occupations of the dense network of eigenstates within an individual subband will serve as efficient channels to assist the exciton transfer. The details, however, will depend on the matrix elements of the density operator. We note that recently it was pointed out that dark states could be efficient in transferring excitations~\cite{dark}.

\begin{figure}
\includegraphics[width=.52\textwidth]{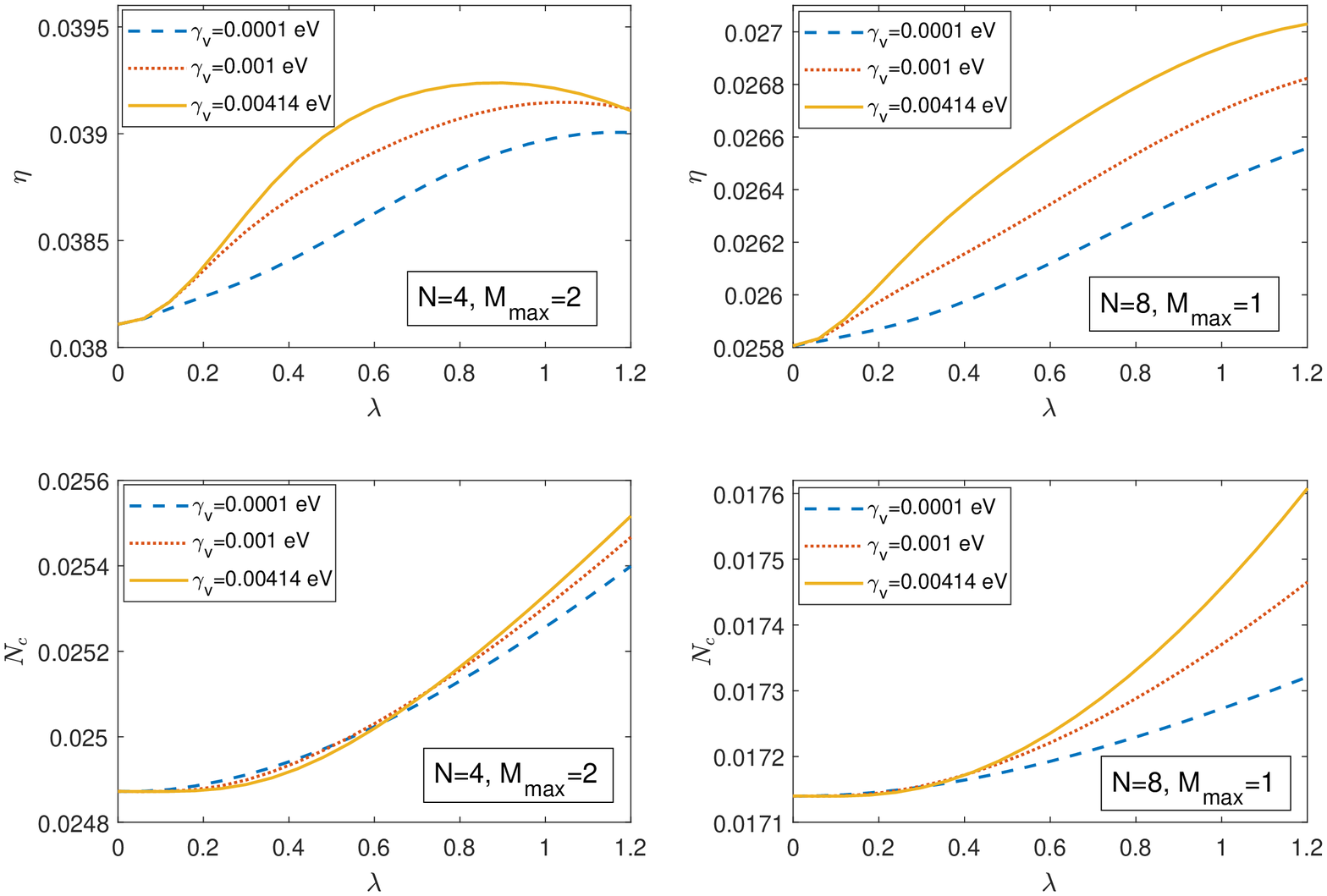}
\caption{Exciton transfer efficiency $\eta$ (upper panels) and the corresponding mean number of photons in the steady state (lower panels) for $N=4$, $M_{\max}=2$ (left column) and $N=8$, $M_{\max}=1$ (right column) at ultrastrong exciton-cavity coupling $g\sqrt{N}/\omega_0=5.0$.
Other parameters: $\varepsilon_n=\omega_{\mathrm{c}}=2$ eV, $\omega_0=0.17$ eV, $J=-\omega_0/4$, $\gamma^{-1}_d=600 ~\mathrm{fs}$, $\gamma^{-1}_p=25 ~\mathrm{fs}$, $\kappa^{-1}=50 ~\mathrm{fs}$, $\gamma^{-1}_L=\gamma^{-1}_R=1 ~\mathrm{ps}$,  and $\bar{n}_L=1$.}
\label{Fig10}
\end{figure}
\par Although the spectrum provides some information on qualitative aspects of the hybrid system, certain steady-state observables are needed to help illustrate the open system dynamics in the long-time limit. To this end, we plot in lower panels of Fig.~\ref{Fig10} the mean number of photons $N_c=\mathrm{Tr}(\rho_{ss} c^\dag c)$ in the steady state for $N=4, M_{\max}=2$ and $N=8, M_{\max}=1$. The corresponding transfer efficiencies are shown in the upper panels. It can be seen that for all cases considered, both $N_c$ and $\eta$ increases with increasing $\lambda$. Thus, the mean number of photons in the steady state can be viewed as a rough measure of the exciton transfer ability. We note that increasing the exciton-cavity coupling also leads to increasing of the photonic part of the lower polaron polariton state~\cite{LPP}, which is consistent with the enhanced transfer efficiency.

\section{Conclusions and Discussions}\label{SecV}
\label{sec-final}
In this work, we studied exciton transport through a source-molecular aggregates-drain setup by treating the exciton-cavity coupling and the exciton-vibration coupling within the molecule on an equal footing. We solve the master equation governing the open dynamics of the exciton-photon-vibration system in truncated vibrational subspaces to obtain the exciton transfer efficiency in the steady state. Starting from investigating the bare exciton-photon system without vibrations, we introduce the definition of exciton transfer efficiency in terms of input and output exciton currents through the molecular chain. Results consistent with previous literatures are obtained, namely the polariton modes formed in the strong exciton-cavity coupling regime dramatically enhance the exciton transfer efficiency.
\par We then turn to study the simultaneous influence of exciton-cavity coupling and exciton-vibration coupling on exciton transfer through molecular chains with different sizes. For a molecular dimer, no vibration-assisted transfer is observed for weak and ultrastrong cavity couplings. However, we do find vibration-assisted exciton transfer for strong cavity coupling at which the Rabi splitting is comparable to the vibration's frequency. For longer molecular chains with $N\geq4$ monomers, we find that the combinational effect of ultrastrong exciton-cavity coupling and strong exciton-vibration coupling result in an enhanced exciton transfer efficiency. Furthermore, it is revealed that finite vibration relaxation could further facilitate the exciton transport in vibrational-enhancement regimes.
\par Although the results presented in this work are based on phenomenological description of dissipative processes and simulations in truncated vibrational spaces, we believe our study provides a preliminary attempt towards understanding vibrational effects on exciton transport in molecular aggregates under strong light-matter interaction. As in Refs.~\cite{FJ2015,LPP,chin}, we employed the rotating wave approximation even in the regime where the collective light-matter coupling is a significant fraction of the bare photon or exciton energies, since multi-excitation manifolds have high detuning with respect to the single-excitation manifold, compared to the Rabi splitting. The counter-rotating terms that induce simultaneous creation of excitonic and photonic excitations are expected to have a non-negligible impact on the exciton transport in the double and higher exciton manifolds.  We leave to future work an analysis on these effects in the ultrastrong coupling regime.

\noindent{\bf Acknowledgements:}
We thank Dazhi Xu for useful discussions and critical reading of the manuscript. This work was supported by the NSFC under Grant Numbers 11705007 and 11891240376, and partially by the Beijing Institute of Technology Research Fund Program for Young Scholars.

\appendix
\section{Equations of motion for a molecular dimer bridge in the absence of vibrations}\label{AppA}
In the zero- and one-excitation subspace spanned by the four basis state $\{|0\rangle=|\mathrm{vac}\rangle, |1\rangle=|1\rangle_{\mathrm{e}}, |2\rangle=|2\rangle_{\mathrm{e}}, |3\rangle= |1\rangle_{\mathrm{c}}\}$, the master equation Eq.~(\ref{ME}) results in
\begin{eqnarray}
\dot{\rho}_{00}&=&[\gamma_d+\gamma_L(1+\bar{n}_L)]\rho_{11}+(\gamma_R+\gamma_d)\rho_{22}\nonumber\\
&&+\kappa\rho_{33}- \gamma_L\bar{n}_L \rho_{00},\\
\dot{\rho}_{11}&=&iJ(\rho_{12}-\rho^*_{12})+ig(\rho_{13}-\rho^*_{13})+\gamma_L\bar{n}_L\rho_{00}\nonumber\\
&&-[\gamma_d+\gamma_L(1+\bar{n}_L)]\rho_{11},\\
\label{rho22}
\dot{\rho}_{22}&=&-iJ(\rho_{12}-\rho^*_{12})+ig(\rho_{23}-\rho^*_{23})\nonumber\\
&&-(\gamma_d+\gamma_R)\rho_{22},\\
\dot{\rho}_{33}&=&-\kappa\rho_{33}-ig(\rho_{13}-\rho^*_{13})-ig(\rho_{23}-\rho^*_{23}),
\end{eqnarray}
for the diagonal elements $\rho_{ii}=\langle i|\rho|i\rangle$ ($i=0,1,2,3$) of $\rho$. It is easy to see that $\frac{d}{dt}\sum^3_{i=0}\rho_{ii}=0$, implying the conservation of the total probability. Note that the dynamics of populations is coupled to that of the coherence $\rho_{12},~\rho_{13}$, and $\rho_{23}$, i.e.,
\begin{eqnarray}
\dot{\rho}_{12}&=&-i\varepsilon_{12}\rho_{12}+ig(\rho_{13}-\rho^*_{23})+iJ(\rho_{11}-\rho_{22})\nonumber\\
&&-\left[\frac{\gamma_R+\gamma_L(1+\bar{n}_L)}{2}+(\gamma_d+\gamma_p)\right]\rho_{12},\nonumber\\
\dot{\rho}_{13}&=&-i(\varepsilon_1-\omega_c)\rho_{13}+ig(\rho_{11}+\rho_{12}-\rho_{33})-iJ\rho_{23}\nonumber\\
&&-\frac{\gamma_L(1+\bar{n}_L)+(\gamma_d+\gamma_p+\kappa)}{2}\rho_{13},\\
\dot{\rho}_{23}&=&-i(\varepsilon_2-\omega_c)\rho_{23}+ig(\rho_{22}+\rho^*_{12}-\rho_{33})-iJ\rho_{13}\nonumber\\
&&-\frac{\gamma_R+(\gamma_d+\gamma_p+\kappa)}{2}\rho_{23},
\end{eqnarray}
where $\varepsilon_{12}=\varepsilon_{1}-\varepsilon_{2}$ is the energy difference between the two monomers.


\begin{thebibliography}{99}
\bibitem{Photosy1} Y.-C. Cheng and G. R. Fleming, Annu. Rev. Phys. Chem. \textbf{60}, 241 (2009).
\bibitem{Photosy2} G. D. Scholes, G. R. Fleming, A. Olaya-Castro, and R. van Grondelle, Nat. Chem. \textbf{3}, 763 (2011).
\bibitem{orgsem1} S. R. Forrest, Nature (London) \textbf{428}, 911 (2004).
\bibitem{orgsem2} S. M. Menke, W. A. Luhman, and R. J. Holmes, Nat. Mater. \textbf{12}, 152 (2013).
\bibitem{PRX} D. J. Gorman, B. Hermmerling, E. Megidish, S. A. Moeller, P. Schindler, M. Sarovar, and H. Haeffner, Phys. Rev. X \textbf{8}, 011038 (2018).
\bibitem{Natm2015} E. Orgiu, J. George, J. Hutchison, E. Devaux, J. F. Dayen, B. Doudin, F. Stellacci, C. Genet, P. Samor\`i, and T. W. Ebbesen, Nat. Mater. \textbf{14}, 1123 (2015).
\bibitem{FJ2015} J. Feist and F. J. Garcia-Vidal, Phys. Rev. Lett. \textbf{114},196402 (2015).
\bibitem{enhanced2015} J. Schachenmayer, C. Genes, E. Tignone, and G. Pupillo, Phys. Rev. Lett. \textbf{114}, 196403 (2015).
\bibitem{Zhou2016} J. Yuen-Zhou, S. K. Saikin, T. Zhu, M. Onbalsi, C. Ross, V. Bulovic, and M. Baldo, Nat. Comm. \textbf{7}, 11783 (2016).
\bibitem{enhanced2017} D. Hagenm\"uller, J. Schachenmayer, S. Sch\"utz, C. Genes, and G. Pupillo, Phys. Rev. Lett. \textbf{119}, 223601 (2017).
\bibitem{Zhong} X. Zhong, T. Chervy, L. Zhang, A. Thomas, J. George, C. Genet, J. A. Hutchison, T. W. Ebbesen, Angew. Chemie \textbf{56}, 9034 (2017).
\bibitem{Zhou2018} M. Du, L. A. Mart\'inez-Mart\'nez, R. F. Ribeiro, Z. Hu, V. M. Menon, and J. Yuen-Zhou, Chem. Sci. \textbf{9}, 6659 (2018).
\bibitem{FJ2018} R. S\'aez-Bl\'azquez, J. Feist, A. I. Fern\'andez-Dom\'inguez, and F. J. Garc\'ia-Vidal, Phys. Rev. B \textbf{97}, 241407 (2018).
\bibitem{SR2018} M. Reitz, F. Mineo, and C. Genes, Sci. Rep. \textbf{8}, 9050 (2018).
\bibitem{Mazza} P. Michetti, L. Mazza, and G. C. La Rocca, in \emph{Organic Nanophotonics} (Springer, Berlin, Heidelberg, 2015), Vol. 39.
\bibitem{Ephy} J. A. \'Cwik, S. Reja, P. B. Littlewood, and J. Keeling, Europhys. Lett. \textbf{105}, 47009 (2014).
\bibitem{Spano2015} F. C. Spano, J. Chem. Phys. \textbf{142}, 184707 (2015).
\bibitem{Spano2016} F. Herrera and F. C. Spano, Phys. Rev. Lett. \textbf{116}, 238301 (2016).
\bibitem{FJPRX} J. Galego, F. J. Garcia-Vidal, and J. Feist, Phys. Rev. X \textbf{5}, 041022 (2015).
\bibitem{LPP} N. Wu, J. Feist, and F. J. Garcia-Vidal, Phys. Rev. B \textbf{94}, 195409 (2016).
\bibitem{KeelingACS} M. A. Zeb, P. G. Kirton, and J. Keeling, ACS Photonics \textbf{5}, 249 (2017).
\bibitem{PRB2005} J. K. Viljas, J. C. Cuevas, F. Pauly, and M. H\"afner, Phys. Rev. B \textbf{72}, 245415 (2005).
\bibitem{PRB2011} A. Yar, A. Donarini, S. Koller, and M. Grifoni, Phys. Rev. B \textbf{84}, 115432 (2011).
\bibitem{Natphy2013} A. W. Chin, J. Prior, R. Rosenbach, F. Caycedo-Soler, S. F. Huelga, and M. B. Plenio, Nat. Phys. \textbf{9}, 113 (2013).
\bibitem{NC2014} E. J. O'Reilly and A. Olaya-Castro, Nat. Comm. \textbf{5}, 3012 (2014).
\bibitem{Plenio2015} N. Killoran, S. F. Huelga, and M. B. Plenio, J. Chem. Phys. \textbf{143}, 155102 (2015).
\bibitem{Datta} S. Datta, \emph{Electronic Transport in Mesoscopic Systems} (Cambridge University Press, 1995).
\bibitem{Guan2013} C. Guan, N. Wu, and Y. Zhao, J. Chem. Phys. \textbf{138}, 115102 (2013).
\bibitem{KeelingnonRWA} J. A. \'Cwik, P. Kirton, S. De Liberato, and J. Keeling, Phys. Rev. A \textbf{93}, 033840 (2016).
\bibitem{Wu2018} N. Wu, Phys. Rev. B \textbf{97}, 014301 (2018).
\bibitem{PRA2014} N. Wu, A. Nanduri, and H. Rabitz, Phys. Rev. A \textbf{89}, 062105 (2014).
\bibitem{PRA1996} R. Houdr\'e, R. P. Stanley, and M. Ilegems, Phys. Rev. A \textbf{53}, 2711 (1996).
\bibitem{Zhang} Q. Zhang, V. Romero-Rochin, and R. Silbey, Phys. Rev. A \textbf{38}, 6409 (1988).
\bibitem{Vidm} L. Vidmar, J. Bon$\rm{\check{c}}$a, M. Mierzejewski, P. Prelov$\rm{\check{s}}$ek, and S. A. Trugman, Phys. Rev. B \textbf{83}, 134301 (2011).
\bibitem{chin} J. del Pino, F. A. Y. N. Schr\"oder, A. W. Chin, J. Feist, F. J. Garcia-Vidal, Phys. Rev. Lett. \textbf{121}, 227401 (2018).
\bibitem{Torma} P. T\"orm\"a and W. L. Barnes, Rep. Prog. Phys. \textbf{78}, 013901 (2015).
\bibitem{dark} C. Gonzalez-Ballestero, J. Feist, E. G. Bad\'a, E. Moreno, and F. J. Garcia-Vidal, Phys. Rev. Lett. \textbf{117}, 156402 (2016).


\end{thebibliography}
\end{document}